\documentclass[12pt, a4paper]{article}
\usepackage[latin1]{inputenc}
\usepackage[dvips]{graphicx,color}
\usepackage{amsmath}
\usepackage{amsfonts}
\usepackage{amssymb}
\numberwithin{equation}{section}
\begin{document}

\newcommand{\be}{\begin{equation}}
\newcommand{\ee}{\end{equation}}

\author{C.~D.~Fosco$^{a}$, G.~Torroba$^{b}$
and H.~Neuberger$^{b}$\\ [7mm]
  {\normalsize\it $^a$Centro At\'omico Bariloche and Instituto Balseiro}\\
  {\normalsize\it Comisi\'on Nacional de Energ\'{\i}a At\'omica}
  \\
  {\normalsize\it R8402AGP Bariloche, Argentina.}\\
  {\normalsize\it $^b$Department of Physics and Astronomy, Rutgers University}\\
  {\normalsize\it Piscataway, NJ 08855, U.S.A} }

\title{\bf \LARGE A simple derivation of the Overlap Dirac Operator}
\maketitle \vskip 1.5cm

\abstract {\noindent We derive the vector-like
four dimensional overlap Dirac
operator starting from a five dimensional Dirac action
in the presence of a delta-function
space-time defect. The effective operator is obtained by first
integrating out all the fermionic modes in the fixed gauge
background, and then identifying the contribution from the
localized modes as the determinant of an operator in one dimension
less. We define physically relevant degrees of freedom on the
defect by introducing an auxiliary defect-bound fermion field and integrating out
the original five dimensional bulk fields.}


\bigskip
\newpage

\tableofcontents

\vskip 1cm

\section{Introduction}

The overlap Dirac operator~\cite{Neuberger:1997fp,Neuberger:1998wv} has a
somewhat unusual form. On the lattice, its determinant gives the exact
effective action that is associated with a pair of left handed fermions in
conjugate representations located on two infinitely separated, parallel
domain walls, embedded in five dimensions, and subjected to the same four
dimensional gauge field background~\cite{Kaplan:1992bt}. The five dimensional gauge field is
independent of the coordinate orthogonal to the domain walls and has no
component in that (fifth) direction~\cite{Kaplan:1992sg,Narayanan:1994gw,Shamir:1993zy}.
Thus, the five dimensional gauge field
is defined by a four dimensional one. The five dimensional space is
$R\times{\cal M}$ where ${\cal M}$ is four dimensional, and for
definiteness, compact and latticized (hence, IR and UV regularized).

The exact effective action is defined by a limiting procedure whereby $R$
is replaced by a finite length segment $L$, imposing specific boundary
conditions at the endpoints of $L$.  A ``bulk'' action is defined from the
system $S\times{\cal M}$ where $S$ is a circle, which is of the same length
as $L$ but now one has translational invariance along $S$. The effective
action is then obtained by dividing the fermion determinant associated with
$R\times{\cal M}$ by the fermion determinant corresponding to the
$S\times{\cal M}$ system, and taking the length of $L$ to infinity
~\cite{Neuberger:1997bg}.

The original derivation was different~\cite{Neuberger:1997fp}, since it dealt first with a
single domain wall embedded in five dimensions, which produced a
formula for the action associated with a single left handed
fermion.  The partition function for the left handed fermion had
an incompletely defined phase, and had to be viewed as a section
of a line bundle over the space of four dimensional gauge fields.
The product of this section with its conjugate produced the
determinant of the overlap operator, a real valued functional of
the four dimensional gauge field which is invariant under gauge
transformations.

In either of the two derivations, the chiral components of the
Dirac fermion live on two separate copies of ${\cal M}$. Although
there is no obstacle to the eventual identification of the two
copies, physically, it would be more appealing to have
the two Weyl components living on the same ${\cal M}$ from the
beginning.

In this note we present a non-rigorous derivation of the overlap formula,
from a starting point where the two chiral components of the Dirac field
live on the same four dimensional manifold ${\cal M}$.  We say that the
derivation is non-rigorous because we do not employ an
explicit UV regularization, although ${\cal M}$ can be taken to be compact.
As we shall see, this approach makes it possible to understand the formula
for the overlap Dirac operator in just a few lines.

\section{The setup}

Thinking in terms of the setup reviewed at the beginning of this note, we
wish to replace the segment $L$ with the two Weyl fields living at its
opposite endpoints, by first gluing the two ends to each other, then
cutting $L$ in the middle, and finally letting the two (new) endpoints
created by the cutting  go to $\pm\infty$, respectively. So, we again have a
$R\times{\cal M}$ manifold.  Denoting the coordinate along $R$ by $s$, we
assume that $s=0$ corresponds to the point where the original endpoints of
$L$ have been glued. The coordinates along ${\cal M}$ are denoted
collectively by $x$. To keep the discussion more general,
we replace the dimension $4$ of ${\cal M}$ by an arbitrary even number $d$.

We must have a singularity, or a defect, at $s=0$, since the two Weyl
fields cannot, in general, be glued continuously to each other.
Nevertheless, the massless Dirac fermion is expected to be bound to that
singularity, i.e., to be localized on the single copy of ${\cal M}$ at
$s=0$.
The defect can be realized by adding a suitable singular mass term to
the $d+1$ dimensional Dirac operator ${\cal D}_{d+1}$, so that the
eigenfunctions of ${\cal D}_{d+1} ~ $ have a discontinuity at $s=0$.

The dimension of $s$ is fixed by the choice of the structure of
${\cal D}_{d+1}$, where first-order derivatives in $s$ and $x$
appear.
Moreover, since the singular mass term must have support at $s=0$,
we conclude that a mass parameter of the form $\xi
\delta(s)$, with a dimensionless parameter $\xi$, is of the right
form. For  ${\cal D}_{d+1}$ to have an eigenvalue problem, we must
assign a value to $\Psi(x,0)$, depending linearly on $\Psi(x,0-)$
and $\Psi(x,0+)$.  To preserve the $s \to -s$ symmetry, which is
tied to four dimensional charge conjugation invariance,
we set $\Psi(x,0)=\frac{1}{2} [ \Psi(x,0-) + \Psi (x,0+)]$.

The  $d+1$ dimensional operator has thus the form:
\begin{equation}
{\cal D}_{d+1}=\gamma_s\partial_s + \not \!\! D + m +\xi\delta(s)
\end{equation}
where $\not \!\!\! D = \not \! \partial + \not \!\!\! A$ is the
$d$-dimensional Dirac operator in the presence of a gauge field
whose components $A_\mu$ depend only on $x$
and our Lie algebra generators are such
that $A_\mu (x) =-A_\mu(x)^\dagger$. We now can determine
$\xi$ by ensuring that ${\cal D}_{d+1}$ has exact zero modes when
the $d$-dimensional gauge field on ${\cal M}$ is from a nontrivial
bundle and as a consequence, for the chosen background, $\not \!\!
D$ has chiral zero modes $\psi_d(x)$.  We wish to ensure that
there exist then zero modes of  ${\cal D}_{d+1}$ of the form
$\Psi(x,s)=\phi(s)\psi_d (x)$. We find that the choice $\xi=-2
\,{\rm sign}(m)$ does the job, with a discontinuous $\phi(s)$
which is zero on the $s>0$ or $s<0$ halves of $R$. On the other
half it goes as $\exp(-|ms|)$. For definiteness, we pick $m>0$ and
$\xi=-2$.

\section{Evaluation of the effective action}

We now evaluate $\Gamma_\xi(A)$, the effective action resulting
from the functional integral corresponding to the action $S_f =
\int d^dx ds \,\bar\Psi{\cal D}_{d+1} \Psi$. We separate the
singular part out, writing ${\cal D}_{d+1}={\mathcal D} +M(s)$
with $M(s)=\xi\delta(s)$. The path integral
\begin{equation}\label{eq:defgxi}
e^{-\Gamma_\xi (A)} \;=\; \int {\mathcal D}\Psi {\mathcal
D}{\bar\Psi}\; e^{-S_f({\bar\Psi},\Psi;\,A)} \;,
\end{equation}
gives the fermion determinant:
\begin{equation}
 {\rm det}_{d+1} \big[ {\mathcal D} + M(s) \big] \;\equiv \; e^{- \Gamma_\xi
 (A)}\;.
 \end{equation}
 Hence:
\begin{equation}
\Gamma_\xi (A) \;=\; - \ln {\rm \det}_{d+1} \big[ {\mathcal D} + M(s) \big]
\;=\; - {\rm Tr}_{d+1} \ln \big[ {\mathcal D} + M(s) \big]\;.
\end{equation}

$\Gamma_\xi$ does not vanish  in the absence of a defect, i.e. when $\xi=0$ and all
modes are de-localized in the $s$ direction.
In order to assign an effective Dirac
operator ${\mathcal O}$ to the additional
modes showing up when $\xi\ne 0$, which are localized at the defect, we define the
effective action $\Gamma_{\mathcal O}$ by subtracting
$\Gamma_0$ from $\Gamma_\xi$:
\begin{eqnarray}
\Gamma_{\mathcal O}(A) &\equiv& \Gamma_\xi (A) - \Gamma_0 (A) \nonumber\\
&=& - {\rm Tr} \ln \big[ I + {\mathcal D}^{-1} M(s) \big] \;,
\end{eqnarray}
where the trace is over all of the spacetime coordinates and indices.

The $\xi = 0$ piece is linear in $L$, the length of the $s$
coordinate, and vanishes for $L=0$. This reflects translational
invariance at $\xi=0$. It is given by:
\begin{equation}
\Gamma_0 (A)\;=\; -L\, \Big[ {\rm tr}(1) ( c_0 \;+\; c_1 \, \ln
\Lambda ) \Lambda \;+\; \frac{1}{2} {\rm Tr}_d (\sqrt{ -\not\!\!
D^2 + m^2 }) \Big],
\end{equation}
where $\Lambda$ is a UV cutoff, $c_1$ and $c_2$ are constants, and
${\rm Tr}_d$ denotes the Dirac and functional trace in
$d$-dimensional space.  The symbol `${\rm tr}$' denotes the Dirac
trace. The $L$ dependence of $\Gamma_0$ makes it possible to
separate $\Gamma_{\mathcal O}$ out from $\Gamma_\xi$.

\subsection{Perturbative derivation}\label{ssec:pert1}
We first calculate $\Gamma_{\mathcal O}$ as a series in powers of $\xi$:
 \begin{equation} \label{eq:power1}
  \Gamma_{\mathcal O} (A)\;=\;\sum_{n=1}^\infty \, \Gamma_{\mathcal O}^{(n)}(A)
 \end{equation}
where:
\begin{equation}
\Gamma_{\mathcal O}^{(n)} (A) \;=\; \frac{(-1)^n}{n} \, {\rm Tr}_{d+1}\left[ \Big(
{\mathcal D}^{-1} \; \hat M \Big)^n \right] \;.
\end{equation}
The operator $\hat M$ has the matrix elements:
$$
<s,x| \hat M | s^\prime , x^\prime >= \xi \delta^d (x-x^\prime)
\delta(s)\delta(s^\prime )\,
$$
Therefore, the integration over the $s$ coordinates can be carried out
explicitly, leaving us with
\begin{equation}\label{eq:power2}
\Gamma_{\mathcal O}^{(n)} (A) \;=\; \frac{(-1)^n}{n} \, \xi^n \;
{\rm Tr}_{d} \left[ {\mathcal K}^n \right] \;,
\end{equation}
with a trace over internal indices and over the $x$ coordinates.
${\mathcal K}$ is the operator that results from evaluating the kernel of
${\mathcal D}^{-1}$ from $s =0$ to $s=0$:
\begin{equation}\label{eq:defk}
    {\mathcal K} \;=\; \int_{-\infty}^{+\infty} \frac{dk_s}{2\pi}
    \; \frac{1}{i \gamma_s k_s + \not \!\! D + m } \;.
  \end{equation}

Writing
$$
 {\mathcal K} \;=\; \int_{-\infty}^{+\infty} \frac{dk_s}{2\pi}
    \;
\left(\frac{- i \gamma_s k_s - \not \!\! D + m}{k_s^2 - \not \!\!
D^2 +
    m^2}\right)\;\;\;\;\;\;\;\;\;\;\;\;\;\;\;
$$
\begin{equation}\label{eq:kr1}
    =\; ( - \not \!\! D + m ) \;
\int_{-\infty}^{+\infty} \frac{dk_s}{2\pi}
    \; \frac{1}{k_s^2 - \not \!\! D^2 + m^2}
    \;,
\end{equation}
we then perform integral over $k_s$, after having thrown away the
piece odd in $k_s$, to obtain:
\begin{equation}\label{eq:kr2}
    {\mathcal K} \;=\; \frac{1}{2} V \;,
\end{equation}
where $V$ denotes the unitary operator:
\begin{equation}\label{eq:defvop}
   V  \;=\; \frac{- \not \!\! D + m }{\sqrt{- \not \!\! D^2 + m^2}}\;.
\end{equation}
Inserting this into the expression for $\Gamma_{\mathcal O}$, we see that:
\begin{equation}\label{eq:gfr1}
\Gamma_{\mathcal O} (A) \;=\; -\; {\rm Tr}_d \left[
\sum_{n=1}^\infty \; \frac{(-1)^{n-1}}{n} \; (\xi \mathcal K)^n
\right] \;,
\end{equation}
or
\begin{equation}\label{eq:gfr2}
\Gamma_{\mathcal O} (A) \,=\,
-\,{\rm Tr}_d \ln \left[ 1 +  \frac{\xi}{2} \;  V \right] \;=\; -\; \ln
{\rm det}_d \left[ 1 + \frac{\xi}{2} \;  V \right] \;.
\end{equation}
We now identify the effective $d$-dimensional operator ${\mathcal
O}$:
\begin{equation}\label{eq:gfr4}
e^{- \Gamma_{\mathcal O}(A)} \;\equiv\; {\rm det}_d {\mathcal O}
\end{equation}
where
\begin{equation}\label{eq:defox}
{\mathcal O} \;=\;  1 + \frac{\xi}{2} \; V  \;.
\end{equation}

With $\xi=-2$ (\ref{eq:defox}) reduces to the usual overlap
operator~\cite{Neuberger:1997fp}:
\begin{equation}
{\mathcal O} \;=\;  1 \,- \, V \;.
\end{equation}
For a trivial gauge field on ${\cal M}=R^4$ and
for small momenta, with $m>0$, \mbox{$V \simeq 1 - \not \!\! D / m$}, leading to
\mbox{${\mathcal O} \; \simeq \; \not \!\! D / m$}.

One could worry about the validity of the expansion
(\ref{eq:power2}) in powers of $\xi$, given that at the end we set
$\xi = - 2$. We see that the expansion really is in the
combination $\xi \mathcal K$ which is
unitary and convergence is not a problem except for eigenstates of $V$
with eigenvalue 1, where we know we should get the logarithm of zero.

\subsection{Auxiliary field derivation}\label{sec:auxf}

To gain further insight into the nature of this system, we now
provide an alternative derivation, which makes use of auxiliary
fields to take the defect into account. These auxiliary fields are
localized at the defect (by definition), and together with
$\Psi(x,0)$ they can be used to make up two linear combinations of
fields on ${\cal M}$: one which has a chirally symmetric
propagator and another one which does not propagate at all. Those
fields do mix, and this is how one can consistently maintain exact
chirality for ``valence'' fermions but not for ``sea'' fermions
which are governed by ${\mathcal O}$, which does not anti-commute
with $\gamma_s$. This is consistent, because ${\mathcal O}$ obeys
the Ginsparg-Wilson relation ~\cite{gw}, which says that the
propagator $\frac{1}{1-V}$ is chiral up to a contact term on
${\cal M}$.

The auxiliary fields arise naturally when we linearize the
singular part of the action, by introducing Grassmann-valued
(Dirac) fields $\chi$ and ${\bar\chi}$~\footnote{We follow the
conventions of \cite{zinn}.}, living in $d$ dimensions:
\begin{equation}\label{eq:aux1}
e^{-\int d^d x ds {\bar\Psi} M(s) \Psi} \;=\; \frac{\int {\mathcal
D}\chi \, {\mathcal D}{\bar\chi} \; e^{- \int d^dx \big(
\frac{1}{\xi} {\bar\chi}(x) \chi(x) - i[{\bar\chi}(x)
  \Psi (x,0) + {\bar\Psi}(x,0) \chi(x) ] \big) }}{\int {\mathcal D}\chi \,
{\mathcal D}{\bar\chi} \;
e^{- \int d^dx \frac{1}{\xi} {\bar\chi}(x) \chi(x)} }
\;.
\end{equation}
We have
\begin{equation}\label{eq:aux2}
e^{-\Gamma_{\mathcal O}(A)} \;=\; \int {\mathcal D}\chi \, {\mathcal
D}{\bar\chi} \; e^{- S_d ({\bar\chi},\chi;A)}
\end{equation}
where:
\begin{eqnarray}
e^{-S_d ({\bar\chi},\chi;A)} & = & {\mathcal N}_\xi \; \big( \det {\mathcal D}
\big)^{-1} \; \int {\mathcal D}\Psi \, {\mathcal D}{\bar\Psi} \; e^{-\int d^dx
\,ds {\bar\Psi} {\mathcal D} \Psi }\nonumber\\
&\times & e^{i \int d^d x  \big[ {\bar\chi}(x) \Psi (x,0) + {\bar\Psi}(x,0)
\chi(x) \big]} \times e^{-\frac{1}{\xi} \int d^d x \bar\chi\chi}
\end{eqnarray}
and ${\mathcal N}_\xi \equiv \det\xi $.
We now integrate over the original Dirac field:
\begin{equation}\label{eq:chiact}
e^{- S_d ({\bar\chi},\chi;A)} \;=\;
e^{-\int d^d x \int d^d x' {\bar \chi}(x) {\mathcal O} (x,x') \chi(x')}\;,
\end{equation}
where
\begin{equation}
{\mathcal O} \;\equiv\;  I \,+\, \frac{\xi}{2} V \;,
\end{equation}
with $V$ as defined in (\ref{eq:defvop}); $\mathcal N _\xi$ was
dropped. We see that:
\begin{equation}
e^{-\Gamma_{\mathcal O} (A)} \;=\;  {\rm det}_d \,{\mathcal O} \;,
\end{equation}
where ${\mathcal O}$ becomes the overlap Dirac operator when
\mbox{$\xi \equiv - 2$}, exactly as in the perturbative derivation.

\section{Degrees of freedom at the defect}

The propagator of the auxiliary fields,
\begin{equation}
\langle {\chi}(x) {\bar\chi}(y) \rangle \;=\; {\mathcal
O}^{-1}(x,y) \;\equiv\; \langle x | {\mathcal O}^{-1} | y \rangle\;,
\end{equation}
has a pole at zero  momentum when $\xi=-2$, describing a
massless dynamical Dirac fermion.  On the other hand, the original
fields, $\Psi (x,0)$ and $\bar\Psi (x,0)$, have the propagator:
\begin{equation}
\langle \Psi (x,0) {\bar\Psi}(y,0) \rangle \;=\;
\langle x |\frac{V}{2 (1 - V) } | y \rangle \;.
\end{equation}

We now form linear combinations of $\Psi(x,0)$ and $\chi(x)$ of
the form $\Phi_\alpha(x)=\Psi(x,0)-\imath \alpha \chi(x)$ and the same
combinations for the independent
fields $\bar\Phi_\alpha(x)=\bar\Psi(x,0)-\imath \alpha \bar\chi(x)$. We find
that for $\alpha=-\frac{1\pm {\sqrt{2}}}{2}~$, $\Phi_\alpha$  has
a propagator proportional to $\frac{1-V}{1+V}$ which anticommutes
with $\gamma_s$ on account of $\gamma_s V \gamma_s=V^\dagger$. On
the other hand, for $\alpha=\frac{1}{2}~$ $\Phi_\alpha$ has a
propagator proportional to $\delta^d (x-y)$, so is non dynamical.
The various $\Phi_\alpha$ fields mix. Physically, on the defect
there is an ordinary massless fermion and also a non-propagating
fermionic degree of freedom that mixes with it.

\section{Conclusion}

The form of the overlap operator in $d$ dimensions and the essential
features that allow the preservation of exact chiral symmetry in the
presence of a UV regulator are understandable with relative ease by
considering states bound to a delta-function mass defect in $d+1$
dimensions, with a particular strength and sign. The
$\delta$-function mass defect plays the role
of the domain walls in Kaplan's~\cite{Kaplan:1992bt} original formulation,
each domain wall being that introduced by
Callan and Harvey for Weyl fermions~\cite{callan}.
The domain walls can be brought infinitesimally close to each other because
the mass parameter on the short segment separating them is taken to infinity.

\vskip 4mm

\subsection*{Acknowledgments}
C.~D.~F acknowledges support by ANPCyT and CONICET, under grants
PICT 17-20434 and PIP 5231, respectively. G.~T. is supported as a
Research Assistant in the Rutgers Department of Physics.
H. N. acknowledges partial support
by the DOE under grant number
DE-FG02-01ER41165 at Rutgers University.

\end{document}